\documentstyle[a4,11pt]{article}
\def\slash{\!\!\!/}
\begin{document}
\title{\huge{Weak decays of doubly heavy hadrons}
\thanks{Research partially supported by CICYT under contract AEN 93-0234
and by IVEI}}
\author{{\bf Miguel Angel Sanchis-Lozano} \\
\\
\it Departamento de F\'{\i}sica Te\'orica and Instituto de F\'{\i}sica
 Corpuscular (IFIC) \\
\it Centro Mixto Universidad de Valencia-CSIC \\
\it Dr. Moliner 50, E-46100 Burjassot, Valencia (Spain) }
\maketitle
\abstract{We explore the application and usefulness of the heavy quark
symmetry to describe the weak decays of hadrons (mesons and baryons)
containing two heavy quarks. Firstly, we address the internal dynamics
of a heavy-heavy bound system with the help of estimates based on potential
models, showing an approximate spin symmetry in the preasymptotic quark mass
region including charmonium, bottonium and $B_c$ meson states. However, no
asymptotic spin symmetry should hold in the infinite quark mass limit
in contrast to singly heavy hadrons. Predictions on semileptonic and
two-body nonleptonic decays of $B_c$ mesons are shown. Furthermore, the
stemming flavor and spin symmetries from the interaction between the
heavy and light components in hadrons (combining in a $\lq\lq$superflavor"
symmetry) permit their classification in
$meson$-type supermultiplets containing singly heavy mesons together with
doubly heavy baryons, and $baryon$-type supermultiplets containing singly
heavy baryons together with some exotic doubly heavy multiquark states
(diquonia). Exploiting their well-defined transformation properties under the
superflavor symmetry group, we get predictions on the widths for some
semileptonic and two-body nonleptonic decays of baryons containing both
$b$ and $c$ \vspace{0.5in} quarks.\newline
\begin{center}
{\it To appear in Nuclear Physics B}
\end{center}
\vspace{1.4in}
\begin{flushright}
FTUV: 94-45
\end{flushright}
\begin{flushright}
IFIC: 94-40
\end{flushright}
\vspace{0.5in}
{\em e-mail} : mas@evalvx.ific.uv.es
\newpage
\large{
\section{Introduction}
History of Science teaches us that the discovery of new particles, far from
complicating our world's description, opens new trends and offers a previously
unknown insight into a more general understanding of the laws of Nature.
In fact, a paradigmatic example in chemistry is the elements' periodic table
whose full significance was not apparent until the development of the
knowledge on the subatomic structure of matter and the birth of
quantum mechanics. Analogously in elementary particle physics, the appearance
of new (though shortly living) particle generations suggests or confirms new
symmetries of matter and its fundamental interactions. Strange, charmed and
beautiful hadrons have been good examples of it in the past and certainly
still will be in the \vspace{0.05in} future.
\par
The existence of doubly heavy mesons and baryons ({\em i.e.} containing any
pair
of heavy quarks either $c$ or $b$) can be viewed as a necessary consequence
of the attractive strong interaction among quarks. So far, however, there has
been no real possibility to produce them at an observable rate with existing
accelerators, leaving aside charmonium and bottonium resonance
\vspace{0.05in} families.
\par
Hopefully, this situation should definitely change with the
advent of high-energy, high-luminosity hadron colliders like Tevatron and
mainly LHC. The two by now accepted experiments at LHC (ATLAS and CMS)
include B-physics in their physical objectives and the feasibility of the
detection of doubly heavy hadrons is being currently addressed \cite{note}
\cite{frid} \cite{leb}. Moreover, a third high-precision experiment at LHC
dedicated to the study of Beauty particles could be finally \vspace{0.05in}
approved.\par
Recently the hadroproduction of heavy quarks in general, and of
doubly heavy mesons and baryons in particular, has attracted
a lot of theoretical work \cite{prod}. Conversely,
no such attention has been yet devoted to the analysis of the subsequent
weak decay of these particles especially in the case of baryons, with the
exceptions \cite{whit} \cite{sav0} \cite{masl} to our \vspace{0.05in}
knowledge.\par
At this stage, the purpose of this work is to make estimates on decay rates
of doubly heavy hadrons in order to conduct an experimental search for their
observation in a not so far future. We have focused, in particular, on initial
state hadrons containing one $b$ quark and one $c$ quark. Indeed, in the case
of mesons it represents the only possibility since the top quark would not form
a bound state due to its large mass while for baryons the reason is
an experimental one \cite{note}. We have tried to base our
calculations on model-independent assumptions reducing whenever possible the
input parameters within the framework of the approximations used in this work.
Our final results, though presumably reasonable, ought to be viewed as a
starting point for a more detailed treatment, as well as a checking limit of
more exact \vspace{0.05in} derivations.\par
This paper is organized as follows. In Section 2 we discuss the (partial)
application of the heavy quark effective theory (HQET), primarily developed
for heavy-light (h-l) systems, to the internal dynamics of bound states of two
heavy quarks. In particular, the appearance of a spin symmetry at leading order
(already pointed out elsewhere \cite{jen}-\cite{mannel}) can be
exploited in the analysis of the weak decays of the
$B_c$ mesons. In Section 3 we also study some weak decay channels of baryons
with $b$ and $c$ quarks by means of a $\lq\lq$superflavor" symmetry, obtaining
predictions on decay widths and branching fractions whose numerical values
are the first so far computed, to our knowledge.
\section{Effective Lagrangians for heavy quarks}
In recent times, several attempts have appeared to describe the heavy quark
strong dynamics from more general grounds than particular models by
simplifying the original QCD Lagrangian into a more manageable form.
On the one hand, the non-relativistic QCD (NRQCD) is a systematic
approach to the dynamics of heavy systems as increasing powers of the quark
velocity inside the hadron \cite{eich}. Once reformulated on a
discrete space-time lattice, it constitutes nowadays an
extremely active field of \vspace{0.05in} research \cite{lati}.
\par
On the other hand, the last few years have witnessed a rapid progress
in a new low-energy effective theory of QCD appropriate when dealing
with hadrons consisting of one heavy component along with light degrees
of freedom. Starting from the original Lagrangian ${\cal L}_{QCD}
=\overline{Q}(iD\slash-m_Q)Q$, HQET can be formulated \cite{neub}
according to the following field redefinitions which remove the large
part of the heavy quark momentum:
\begin{eqnarray}
h_v(x) & = & e^{im_Qv{\cdot}x}P_+Q(x) \\
H_v(x) & = & e^{im_Qv{\cdot}x}P_-Q(x)
\end{eqnarray}
where $P_{\pm}=(1{\pm}v{\slash})/2$ denote the usual projectors onto
positive/negative energy states and $h_v$ and $H_v$ stand for the
$\lq\lq$large" and $\lq\lq$small" components of the spinor field. In
order to deal with hadrons containing a heavy
antiquark, one must introduce the corresponding fields $h_v^-$ and
$H_v^-$ \cite{neub}\vspace{0.05in}.\par
The degrees of freedom corresponding to $H_v$ can be eliminated by means
of the equations of motion of QCD yielding a non-local effective
Lagrangian. Then, a $1/m_Q$ power expansion in terms of local operators
is achieved leading to
\begin{equation}
{\cal L_{eff}}^{(i)}\ =\ \overline{h}^{\ (i)}_viv{\cdot}D\ h^{(i)}_v\ +\
\frac{1}{2m_{Q_i}}
[\ {\cal K^{(i)}}+Z(m_{Q_i}/\mu)\ S^{(i)}\ ]\
+\ {\cal O}(1/m_{Q_i}^2)
\end{equation}
where the superscript $i$ labels the heavy flavor and those terms including
\begin{eqnarray}
{\cal K^{(i)}}   & = & \overline{h}^{\ (i)}_v(iD_{\bot})^2h^{(i)}_v \\
{\cal S^{(i)}}   & = & \frac{g_s}{2}\ \overline{h}^{\ (i)}_v
{\sigma}^{\alpha\beta}G_{\alpha\beta}\ h^{(i)}_v
\end{eqnarray}
correspond to the (gauged) kinetic energy of the quark $i$ in the hadron's
rest frame, and the analog of the Pauli term describing the chromomagnetic
interaction of the quark $i$ with the light degrees of freedom. $D_{\bot}$
denotes the transverse covariant derivative, $D_{\bot}=D-v(v{\cdot}D)$. The
renormalization coefficient $Z$ takes into account the short-distance physics
in the chromomagnetic \vspace{0.05in} interaction \cite{neub}.\par
The leading contribution in the Lagrangian of eq. (3) represents
a static quark field, possessing two types of new symmetries not present in
the original theory. In fact, the symmetry of this effective theory is larger
than merely the product of the flavor and spin ones for the fermionic heavy
quark fields, even extending to systems in which the
massive colored particles have different spins. This issue will be
discussed in more detail in Section 2.2 and used in Section 3.
\subsection{Doubly heavy versus singly heavy hadronic systems}
In a weakly bound state of two heavy quarks such as charmonium or
bottonium, the residual momentum $k$ (defined as the quark's momentum
measured in the hadron's rest frame) neither should be too large on the
average, amounting to a relative small off-shellness of the constituents
\vspace{0.05in} \cite{kor}.\par
Indeed, according to a large variety of potential models $<k/m_Q>$
(directly related to the average internal quark velocity $v'$
in non-relativistic motion) scales down with some power of the
quark mass, {\em i.e.} $<k/m_Q>{\sim}\ m_Q^{-\nu}$ where $\nu$ takes a
value near $0.5$ for a logarithmic-like potential, or near $0.34$ for a
$\lq\lq$Cornell-type" potential \cite{kwon}. This behavior must be understood
limited to a certain mass region becoming inappropiate in a
dominant short-distance \vspace{0.05in} regime.
\par
Motivated by those arguments, we shall insist in applying
the inverse heavy quark expansion generated by the field redefinitions
(1) and (2) in order to get an effective non-relativistic Lagrangian to
describe heavy-heavy (h-h) bound systems. As we shall see, important
limitations of the low-energy effective theory will appear with regard to
h-l systems, in particular the lack of flavor symmetry even at leading order
whereas a spin symmetry will hold in a certain preasymptotic mass
\vspace{0.05in} domain.
\par
In h-h systems the internal motion of the heavy constituents cannot
be neglected even at leading order. This can be viewed as a consequence of
the delicate balance between the average potential and kinetic energies
in the bound state, as pointed out a long time ago in terms of the virial
theorem \cite{brow}. Thereby, one must keep the kinetic piece even at lowest
order yielding a Schr\"{o}dinger-type Lagrangian plus the Pauli \vspace{0.05in}
term
\par
\begin{equation}
{\cal L_{eff}^{(i)}}\ =\ {\cal L_0^{(i)}}\ +\
\frac{1}{2m_{Q_i}}\ Z(m_{Q_i}/\mu)\ {\cal S^{(i)}}\ +\ {\cal O}(1/m_{Q_i}^2)
\end{equation}
where
\begin{equation}
{\cal L_0^{(i)}}\ =\
\overline{h}^{\ (i)}_viv{\cdot}D\ h^{(i)}_v\  +\ \frac{{\cal
K}^{(i)}}{2m_{Q_i}}
\end{equation}
The term including the operator (5) describes the chromomagnetic
interaction of the quark $i$ with the gluon field generated by itself and
its heavy \vspace{0.05in} partner.\par
Let us stress that in the Lagrangian expansion for a h-h
system it is the order in the internal velocity $v'$ rather
than the dimension of the operators what matters in neglecting further
terms \footnote{This is the crucial difference with respect to h-l
systems where $v'\ {\simeq}\ {\Lambda}_{QCD}/m_Q$.}. Thus, from arguments
based on power-counting rules \cite{lepa} the expected magnitude of the Pauli
term is suppressed by a power $v'^2$ with respect to ${\cal L_0}$. In
fact, observe that for the sake of consistency in the characteristic
magnitude, one should have considered some {\em further terms} of the
$m_{Q_i}^{-1}$ expansion contributing at the same order in the internal quark
velocity. Such next terms (up to order $v'^4$) would include relativistic
corrections to the kinetic energy as well as spin-dependent effects like
spin-orbit forces. The latter, however, vanish for $s$-wave states so that
in this case the only relevant spin effect comes from the operator (5) while
the former can be incorporated into a spin-independent Lagrangian beyond
\vspace{0.05in} ${\cal L_0^{(i)}}$.\par
Once taken into account, the chromomagnetic operator resolves the degeneracy
between the spin-singlet and spin-triplet states, both in h-l
systems as in h-h systems. An important difference arises, however, in
the behavior of the hyperfine splitting in the $m_Q{\rightarrow}\infty$
limit. In h-l systems the degeneracy becomes asymptotically more and
more exact. In sharp contrast, the validity of an approximate spin symmetry in
doubly heavy states will be justified in a certain intermediate quark
mass region, but it should certainly be broken above a more or less defined
$\lq\lq$threshold". We examine below this issue with the help of potential
models. (Although potential models do not come directly from first principles
they describe very well the general features of hadron \vspace{0.05in}
spectroscopy).\par
In order to assess the relative importance of the Pauli term in (6), we shall
assume a Coulomb-plus-linear potential
\[ V(r)\ =\  V_0\ -\frac{\kappa}{r}\ +\ {\sigma}r \]
For those states above the pure Coulomb well, the wavefunction at the origin
is largely determined by the confining potential. In particular for the
1S state, ${\mid}\psi(0){\mid}^2\ {\approx}\ e_1(\hat{m}_{ij}{\sigma}/2\pi)$
\cite{luch}, where $e_1=1.935$ and $\hat{m}_{ij}=m_{Q_i}m_{Q_{j}}/
(m_{Q_i}+m_{Q_{j}})$ stands for the reduced mass of the $Q_iQ_{j}$ system. For
definiteness and simplicity, we ascribe the non-confining part of the
potential to single-gluon exchange, hence $\kappa=c_F{\alpha}_s$
where $c_F=4/3,\ 2/3$ is a color factor for a quark-antiquark pair,
quark-quark pair respectively. We thus write the well-known expression
\cite{luch} for the hyperfine mass splitting as
\begin{equation}
{\Delta}M{\mid}_{hyp}\ =\ \frac{8c_F\pi}{3}
\frac{{\alpha}_s}{m_{Q_i}m_{Q_{j}}}\ {\mid}\psi(0){\mid}^2
\end{equation}
Inserting the ${\mid}\psi(0){\mid}^2$ expectation
\begin{equation}
{\Delta}M{\mid}_{hyp}\ {\simeq}\ \frac{8c_F}{3}
\frac{{\alpha}_s\sigma}{m_{Q_i}+m_{Q_j}}
\end{equation}
and we see that ${\Delta}M{\mid}_{hyp}$ decreases whenever any $m_{Q_i}$
increases. In the particular case of $J/\psi$ or $\Upsilon$, $m_{Q_i}=m_{Q_j}
=m_Q$ and \vspace{0.05in} ${\Delta}M{\mid}_{hyp}\ {\simeq}\
16{\alpha}_s\sigma/9m_Q$.
\par
One can establish a relationship between the above expressions based on
a potential model and an analogous definition of the ${\lambda}_2$
parameter for h-l mesons \cite{neub} \cite{falk}, now for a doubly heavy bound
system. To this end, we consider the spin-spin effect in the
mass spectrum due to the chromomagnetic interaction of the quark $i$ with
a background of color field according to
\begin{equation}
{\delta}M^{(i)}\ =\ -\ \frac{<H(v){\mid}{\int}d^3x\
{\delta}{\cal L_{eff}^{(i)}}(x){\mid}H(v)>}{<H(v){\mid}H(v)>} \]
where the correction term is here in particular
\[ \delta{\cal L_{eff}^{(i)}}\ =\ \frac{1}{2m_{Q_i}}\
Z(m_{Q_i}/{\mu}_i)\ {\cal S^{(i)}}
\end{equation}
and the hadron states are normalized non-relativistically. Let us
now introduce
\begin{equation}
\frac{<H(v){\mid}{\cal S^{(i)}}(0){\mid}H(v)>}
{<H(v){\mid}\overline{h}_v^{(i)}h_v^{(i)}{\mid}H(v)>}\
=\ d_H\ {\lambda}^{(i)}_2
\end{equation}
where $d_H=3,\ -1$ for the spin-singlet, spin-triplet state respectively.
As far as we will be interested in $s$-wave states, the spin-spin interaction
vanishes everywhere except in the origin $\vec{x}=\vec{0}$. Thereby, taking
into account that with the above normalization the denominator of (11) is
unity, one can readily identify
\vspace{0.05in} ${\delta}M^{(i)}{\mid}_{hyp}=-d_HZ{\lambda}_2^{(i)}/2m_{Q_i}$.
\par
In reality, the hyperfine effect can be $\lq\lq$ascribed" to either heavy
quark $i$ or $j$. Because of the symmetry of the spin interaction
between both heavy quarks in a $s$-wave state (of the dipole-dipole type), the
following identity can be formulated by requiring that
${\delta}M^{(i)}{\mid}_{hyp}={\delta}M^{(j)}{\mid}_{hyp}$:
\begin{equation}
Z(m_{Q_i}/{\mu}_i)\ m_{Q_{j}}{\lambda}_2^{(i)}({\mu}_i)\ =\
Z(m_{Q_{j}}/{\mu}_j)\ m_{Q_i}{\lambda}_2^{(j)}({\mu}_j)\  =\
Z(\hat{m}_{ij}/\mu)\ \hat{m}_{ij}\hat{{\lambda}}_2(\mu)
\end{equation}
where $\hat{\lambda}_2$ may be interpreted as resulting from the
chromomagnetic interaction of a fictitious quark of mass $\hat{m}_{ij}$
moving with the quarks' relative velocity through the background of the
color field generated by both $i$ and $j$ quarks. Now, including a factor
$1/2$ to avoid double counting, we write in an explicitly symmetric form
\[ {\Delta}M{\mid}_{hyp}\ =\ M_{H^{\ast}}-M_{H}\ =\ \frac{1}{2}\
 [{\delta}M^{(i)}{\mid}_{hyp}+{\delta}M^{(j)}{\mid}_{hyp}]\ =  \]
\begin{equation}
\biggl[\ Z(m_{Q_i}/{\mu}_i)\ \frac{1}{m_{Q_i}}
{\lambda}_2^{(i)}({\mu}_i)\ +\ Z(m_{Q_{j}}/{\mu}_j)\ \frac{1}{m_{Q_{j}}}
{\lambda}_2^{(j)}({\mu}_j)\ \biggr]\
=\ Z(\hat{m}_{ij}/\mu)\ \frac{2}{m_{Q_i}+m_{Q_j}}\ \hat{\lambda}_2(\mu)
\end{equation}
and thus $Z\hat{\lambda}=(M_{H^{\ast}}^2-M_H^2)/4$, in analogy to h-l
systems \cite{falk}. By comparison with eq. (9) we further get
\begin{equation}
Z(\hat{m}_{ij}/\mu)\ \hat{\lambda}_2(\mu)\ \simeq\
\frac{4c_F}{3}\ {\alpha}_s(r_0^{-1})\ \sigma
\end{equation}
where $r_0$ stands for the typical radius of the hadron, {\em i.e.}
$r_0^{-1}\ {\simeq}\ c_F{\alpha}_s(r_0^{-1})\hat{m}_{ij}$. Note that
$\hat{\lambda}_2$ is proportional to {\em the string
tension} $\sigma$ and depends on the renormalization scale $\mu$ but
$Z(\mu)\hat{\lambda}_2(\mu)$ does not. Needless to say that eq. (14)
must not be taken literally due its naive model dependence but
rather as an indication of the close relationship between the
$\hat{\lambda}_2$ parameter and the string tension. Thus, the
behavior of the hyperfine mass splitting for the $J/\psi$ and
$\Upsilon$, roughly of the type $m_Q^{-1}$ as in h-l systems, amounts to
the $\hat{m}_{ij}$-independence of \vspace{0.05in} $\hat{\lambda}_2$.
\par
A numerical estimate using ${\alpha}_s\ {\simeq}\ 0.4$,
${\sigma}=0.18$ GeV$^{2}$
yields $\hat{\lambda}_2(\mu{\simeq}1$GeV) ${\simeq}\ 0.13$ GeV$^2$,
in significant accordance with the respective value of
${\lambda}_2(\mu{\simeq}2\overline{\Lambda}$) ${\simeq}\ 0.15$ GeV$^2$
for a h-l meson. Indeed, this is not surprising for it could be
directly guessed from the numerical similarity between the mass splittings
of the  $D$, $D^{\ast}$ states and the ${\eta}_c$, $J/\psi$ states
\vspace{0.05in}.\par
In summary, we advocate the validity of the application of the spin symmetry
to the analysis of the weak decays of doubly heavy hadrons in analogy to
singly heavy hadrons on the basis of the $m_Q$-independence and smallness
of \vspace{0.05in} $\hat{\lambda}_2$.
\par
Nevertheless, this emerging symmetry cannot be considered as
asymptotically valid in h-h states. Since in the Coulomb regime
$v'\ {\sim}\ {\alpha}_s$, we may then write ${\Delta}M{\mid}_{hyp}$ $\sim$
$m_Q{\alpha}_s^4$. The $\lq\lq$transition" mass domain where the
linear behavior takes over the logarithmic fall off can be crudely
estimated (from above) to be around $m_0$ ${\simeq}$ $e^4{\Lambda}_{QCD}$
${\simeq}$ $10$ GeV. Only below it (thereby including $J/\psi$, $B_c$ and
${\Upsilon}$ states) spin symmetry can be used as an
approximation. Indeed, in the language of potential models, a dominant
Coulomb regime determines the wavefunction at the origin as
${\mid}\psi(0){\mid}^2=(c_F{\alpha}_s\hat{m}_{ij})^3/\pi$ leading to the
growth of the hyperfine splitting as follows from eq. (8). Stated in other way,
$\hat{\lambda}_2$ would be proportional to the reduced mass squared yielding
the growth of \vspace{0.05in} ${\Delta}M{\mid}_{hyp}$.\par
\subsection{Superflavor multiplets}
In the preceding Section we analyzed the interaction dynamics of a pair
of heavy quarks borrowing to some extent the formalism developed for
a h-l system. In the following, we shall examine the interaction of a heavy
diquark with a light component on the basis of an $\lq\lq$additional"
effective $diquark$ \vspace{0.05in} theory.\par
Let us start by recalling that QCD has, among others, a notable property:
two color triplets can bind together to yield another color (anti)triplet. In
particular, if both quarks are heavy enough, they will form an almost pointlike
source of color behaving as a heavy antiquark with respect to a third
light component in a hadron. This property of QCD, combined with the
stemming spin-flavor symmetry at the very large quark mass limit should
permit to classify hadrons containing one quark and two heavy antiquarks in
multiplets for the same velocity with similar properties {\em as far as
the light degrees of freedom are concerned}. Indeed,
the fundamental representation of the $SU(6){\otimes}\ U(1)$ spin-flavor
symmetry group consists of the two spin states of the fermionic quark
field, the scalar di-antiquark field and the three spin states of the
axial-vector di-antiquark field. All these objects appear to be the same to
any bound light antiquark \footnote{The underlying superflavor symmetry in the
very heavy mass limit of the strong interaction was firstly developed by
Georgi, Carone, Wise and Savage \cite{geor}\cite{sav}.}. In order to
explicitly exhibit this symmetry, we shall write the representation
of the superflavor group as $\lq\lq$quark" supermultiplets (for each
fixed $v$) expressed as {\em nine}-component column vectors
\begin{center}
\[ {\Psi}_v\ =\ \left( \begin{array}{c}
h_v \\ S_v \\ A_v^{\mu} \end{array} \right) \]
\end{center}
where $A_v^{\mu}$ satisfies the constraint $v_{\mu}A_v^{\mu}=0$. (We have
omitted any reference to the flavor of the constituents). The effective
Lagrangian then reads
\[ {\cal L_{eff}}\ =\ \frac{1}{2}\ \overline{\Psi}_v{\cal M}\ iv{\cdot}
^{^<}\overline{D}^{>}\ {\Psi}_v \]
where ${\cal M}$ is a $9{\times}9$ mass matrix depending on the normalization
of the fields which can be taken from \cite{geor} as
\[ {\cal M}\ =\ \left(
\begin{array}{ccc}
1 & 0 & 0 \\ 0 & 2m_S & 0 \\ 0 & 0 & -2m_A
\end{array} \right) \]
and
\[ \overline{\Psi}_v\ =\ {\Psi}_v^{\dag}\ \left(
\begin{array}{rrr}
{\gamma}^0\ 0\ 0 \\ 0\ 1\ 0 \\ 0\ 0\ g
\end{array} \right) \]
where \vspace{0.05in} $g$ = diag$(1,-1,-1,-1)$.\par
One can form a superflavor $\lq\lq$$meson$" multiplet by the tensor product
of ${\Psi}_v$ and one light antiquark field (there is no need to specify
its flavor assuming unbroken light flavor $SU(3)$). Therefore, such hadronic
supermultiplet puts together singly heavy mesons and doubly heavy
antibaryons. The latter, though with different quantum numbers, really look
like $\lq{\lq}mesons"$ in many respects, as commented more extensively in
Section 3. In the infinite quark mass limit, the behavior of the light degrees
of freedom becomes independent of the heavy constituent flavors $i$, $j$ and of
the di-antiquark spin. Let us note that such supermultiplet is, however, not
completely flavor-independent even at this asymptotic limit because of the
mass dependence of the heavy di-antiquark internal dynamics
\footnote{One may extend the often quoted pedagogical analogy between the
atom and a h-l meson. Now, the physics of the compound nucleus (the
di-antiquark) is not at all the same for different isotopes although they
have similar chemical properties (those of the light degrees of freedom).}. On
the other hand, singly heavy baryons cannot be included in any $meson$
supermultiplet since the groundstate light di-quark (either spin-0 or
spin-1) should behave quite differently than a spin-$1/2$ light
\vspace{0.05in} antiquark.
\par
Instead, one can construct a $baryon$ superflavor multiplet by combining
either one heavy quark or one heavy di-antiquark with a light
di-quark \footnote{Henceforth, the reader himself will kindly realize
whether we are referring to quarks or antiquarks so for the sake of brevity
we generally suppress any explicit reference to antiquarks.}. Moreover, we
sort out two possibilities according to the spin of the light component. If
it is an spinless object the supermultiplet contains ${\Lambda}_Q$ baryons
together with doubly heavy diquonia \footnote{Historically four-quark systems
were in fact referred to as $\lq\lq$baryonia" partly due to their expected
decay
into a baryon-antibaryon pair \cite{barn}. In references \cite{lip}
\cite{zou}  \cite{sema} the similarity between
$Q_iQ_j\overline{q}\ \overline{q}$ states and ${\Lambda}_Q$ baryons
is established in the context of quark/potential models. Let us also mention
that there is an alternative interpretation for the existence of such
exotic states as di-mesons, {\em i.e.} $(Q_i\overline{q})-(Q_j\overline{q})$
states \cite{wis2}. Obviously, the latter would not be the supersymmetric
partners of singly heavy baryons.} ($Q_iQ_j\overline{q}\ \overline{q}$). Other
supermultiplets would include baryon (${\Sigma}_Q,\ {\Sigma}_Q^{\ast}$)
states with a spin-1 light component.\par
\section{Weak decays of hadrons with two heavy quarks}
\par
Hadrons (mesons or baryons) containing two heavy quarks $Q_iQ_j$ are
particularly interesting since their lowest lying states would only decay
weakly in contrast to charmonium and bottonium resonances. In the case of
mesons, the only available possibility of gathering together two
different heavy quarks is the $B_c$ meson. This is probably the reason why
$B_c$ is nowadays object of growing attention both because of
its weak decay and in hadron spectroscopy due to its
interpolating position between the $J/\psi$ and the
$\Upsilon$ \cite{quig}.\par
\subsection{Semileptonic decay of the $B_c$ meson}
\par
As pointed out in the pioneering works by Nussinov and Wetzel
\cite{nussi}, Voloshin and Shifman \cite{shif} the description of
the decay of a h-l hadron like a $B$ meson simplifies dramatically
in the infinite quark mass limit. When the heavy quark decays into another
still massive quark, the surrounding brown-muck acts as a pure spectator
with only one form factor (the Isgur-Wise universal function $\xi(w)$,
$w=v_1{\cdot}v_2$) governing the transition dynamics of the light degrees
of freedom. At zero recoil $\xi(w=1)$ is unity, reflecting a
perfect overlap between the initial and final hadron wave \vspace{0.05in}
functions \cite{neub}.\par
On the other hand, it has been argued in recent literature
about the possibility of testing the spectator behavior by experimentally
discriminating among specific channels for $b$ or $c$ decays \cite{frid}
in $B_c$ ($\overline{b}c$) mesons. Let
us mention, however, that a simple-minded spectator picture of the $B_c$
decay should differ with regard to the analogous decays of $B$ and $D$
mesons. In h-l mesons the spectator behavior makes sense as an asymptotic
expectation whereas both heavy quarks are necessarily
involved in the dynamics of a h-h hadron even for infinite
constituent \vspace{0.05in} masses.
\par
Let us start by considering the $\overline{b}$ quark undergoing the decay,
yielding the exclusive semileptonic channels:\par
\begin{equation}
B_c\ {\rightarrow}\ {\eta}_c(J/\psi)\ +\ \overline{\ell}\ \nu
\end{equation}
Applying the spin symmetry discussed in Section 2 (see the appendix for more
details), the expressions for the widths obtained in the non-recoil
approximation are
\begin{eqnarray}
{\Gamma}(B_c{\rightarrow}{\eta}_c+\overline{\ell}\nu) & {\simeq} &
\frac{G_F^2m_1^5}{192\ {\pi}^3}\ \varphi(x)\ f_{12}^2\ {\mid}V_{cb}{\mid}^2 \\
{\Gamma}(B_c{\rightarrow}J/\psi+\overline{\ell}\nu) &  {\simeq} &
\frac{G_F^2m_1^5}{192\ {\pi}^3}\ [\varphi(x)+\phi(x)]\ f_{12}^2\
{\mid}V_{cb}{\mid}^2
\end{eqnarray}
where $x=m_2/m_1$ and $\varphi$, $\phi$ are phase space factors for vanishing
leptonic masses whose expressions can be found in the appendix. Notice
that we keep both $\varphi$, $\phi$ functions separately to facilitate
later comparisons. The form factor is given by
$f_{12}=\sqrt{m_2/m_1}\ \hat{\eta}_{12}(v_1=v_2)$, where the overlap factor
$\hat{\eta}_{12}$ is not fixed at zero recoil by a current conservation due to
the lack of an underlying flavor symmetry. Instead, it can be estimated by
means of a non-relativistic model of hadrons (see appendix) like the ISGW
model \cite{igsw} according to expression \vspace{0.05in} (A.9).\par
Next we want to apply our simplified approach to the decaying $c$-quark
in the semileptonic channels of $B_c$ mesons:
\begin{equation}
B_c\ {\rightarrow}\ B_s(B_s^{\ast})\ +\ \overline{\ell}\ \nu
\end{equation}
where the heavy quark spin symmetry and the non-recoil approximation
will be employed once again. In \cite{jen} the authors made an estimate
of the overlap integral between the $B_c$'s and $B_s$'s wavefunctions,
denoted by $\hat{\Omega}$, by means of an
operator product expansion retaining only the leading contribution. We have
rederived the expression for $\hat{\Omega}$ at zero recoil resulting, however,
\begin{equation}
\hat{\Omega}\ {\simeq}\
4\ \biggl(\frac{\pi}{3}\biggr)^{1/2}f_{B_s}\sqrt{m_{B_s}}\ r_0^{3/2}
\end{equation}
where $r_0=1/c_F\hat{m}_{bc}{\alpha}_s(\hat{m}_{bc})$ is the radius of the
$B_c$ meson with $\hat{m}_{bc}$ denoting the reduced mass of the $b$ and $c$
quarks.  Let us note some discrepancies between $\lq\lq$our" formula and the
original one presented in \cite{jen}. On the one hand, there is a factor
$\sqrt{2}$ apparently due to a different normalization of states
in (to $2v^0$ in our case instead of $v^0$). Moreover,
there is an extra factor $\sqrt{3}$ in the denominator due to color
\footnote{In writing the initial $B_c$ state, a sum over colors must be
understood in eq. (4.1) of ref. \cite{jen} and a normalization factor
$1/\sqrt{3}$ should be included. Keeping
color indices throughout, {\em i.e.} in the wavefunction and the color
singlet weak current and taking into account the color Kr\"{o}necker
delta implicitly contained in eq. (4.4) of \cite{jen}, the normalization factor
$1/\sqrt{3}$ survives all the steps of the calculation.}. In fact, let us
note that $f_{B_s}\sqrt{m_{B_s}}/2\sqrt{3}$ can be interpreted as the wave
function ${\psi}_{B_s}(0)$ at the origin according to quark models.
Therefore, the $\lq$$\lq$constant" ${\psi}_{B_s}(0)$ factorizes out in the
overlap integral of the initial and final wavefunctions. (This can be
accounted for since the variation of ${\psi}_{B_s}$ takes place over longer
distances than the typical size $r_0$ of the $B_c$). However, let us
observe that the evaluation of $\hat{\Omega}$ in (19) should
be somewhat overestimated because of the radial decrease of the $B_s$
\vspace{0.05in} wavefunction.\par
The decay widths come finally to be
\begin{eqnarray}
{\Gamma}(B_c{\rightarrow}B_s+\overline{\ell}\nu) & {\simeq} &
\frac{G_F^2m_1^5}{192\ {\pi}^3}\ \varphi(x)\ f_{12}^2\ {\mid}V_{cs}{\mid}^2 \\
{\Gamma}(B_c{\rightarrow}B_s^{\ast}+\overline{\ell}\nu) &  {\simeq} &
\frac{G_F^2m_1^5}{192\ {\pi}^3}\ [\varphi(x)+\phi(x)]\ f_{12}^2\
{\mid}V_{cs}{\mid}^2
\end{eqnarray}
where $f_{12}=\sqrt{m_2/m_1}\ \hat{\Omega}$.
\par
Setting usual numerical values in (19) one gets from the above expressions
widths of order $10^{-13}$ GeV, exceedingly larger than those expected from
quark-model calculations, for example \cite{lusi}. We therefore think that
expression (19) actually overestimates the widths in the non-recoil
approximation so we prefer to evaluate $\hat{\Omega}$ employing the ISGW
model according to eq. (A.9) once \vspace{0.05in} again.\par
In Table I we present our estimates turning out to be in remarkable agreement
with the quark model predictions of ref. \cite{lusi} for the same input
parameters. (In fact, our results can be considered as oversimplified
quark model calculations). This agreement extends as well over the results
by other authors in spite of the simplicity of our approach. There
is, however, an important discrepancy (common to all other references quoted
in Table I) with the results by Bagan {\em et al} relative to the widths
of those semileptonic channels with a final vector meson (either a $J/\psi$ or
a $B_s^{\ast}$).
\subsection{Two-body nonleptonic decay of the $B_c$}
In this Section we address some two-body weak decays of the $B_c$ meson. Let
us focus on the process induced by the decay of the $b$-quark,
\begin{equation}
B_c\ {\rightarrow}\ {\eta}_c(J/\psi)\ +\ \pi
\end{equation}
The exclusive channel corresponding to a final $J/\psi$
is one of the most promising from the point of view of its experimental
detectability at a hadron collider \vspace{0.05in} \cite{note}.\par
We shall base our theoretical approach on two hypothesis. The assumption
of the factorization permits to relate such decay with the
semileptonic one at $q^2=m_{\pi}^2\ {\simeq}\ 0$, according to \cite{bjor}
\[ \frac{\Gamma(B_c{\rightarrow}{\eta}_c(J/{\psi})+\pi)}
{d{\Gamma}/dq^2{\mid}_{q^2=0}(B_c{\rightarrow}{\eta}_c(J/{\psi})+
\overline{\ell}{\nu})}\
{\simeq}\ 6{\pi}^2f_{\pi}^2\ {\mid}V_{ud}{\mid}^2\ {\simeq}\ 1\ GeV^2 \]
Besides, we keep the non-recoil approximation in the evaluation
of the hadronic matrix element. Thus, making use of the set of
equations (A.6) and (A.7) a straightforward derivation yields
\begin{equation}
{\Gamma}(B_c{\rightarrow}{\eta}_c(J/{\psi})+\pi)\ {\simeq}\
\frac{G_F^2(m_1^2-m_2^2)^3}
{192\ {\pi}^3m_1^3}\ f_{12}^2\ {\mid}V_{cb}{\mid}^2
\end{equation}
where $m_2$ stands either for the mass of the ${\eta}_c$ or the $J/\psi$. As
seen from Table II, the agreement with other quark model predictions is
remarkable. Furthermore, the general accordance manifested both in tables I
and II suggests the extension of our simplified formalism to the weak decays
of baryons with two heavy quarks.
\subsection{Semileptonic decay of a doubly heavy baryon}
It has been pedagogically introduced in the context of HQET the analogy
between an atom and a meson containing one heavy quark along with light
degrees of freedom \cite{grins}. On the other hand, in baryons with two heavy
quarks ($Q_iQ_jq$), the massive quark pair should bind into a $\overline{3}$
source of color interacting with the light degrees of freedom
as a static antiquark \cite{sav} \cite{ito} \cite{anse}. Assuming
one-gluon exchange, the Bohr radius of the heavy diquark is
$\tilde{r}_0\ {\simeq}\ 1/c_F{\alpha}_s(r_0^{-1})\hat{m}_{ij}\ {\simeq}\ 2r_0$,
where $\hat{m}_{ij}$ is the reduced mass of the ($Q_iQ_j$) subsystem. Indeed,
one expects the size of the heavy diquark be presumably smaller than the
typical length of the hadron, so that the
analogous picture of a doubly heavy baryon in atomic physics would be
a deuterium atom rather than an hydrogen \vspace{0.05in} molecule.\par
If the two heavy quarks have different flavors there are two possible spin
configurations of the compact diquark at lowest level. Supposing that the
$(Q_iQ_j)$ forms a spin-singlet (thus flavor-singlet)
state the ground-state hadron will have spin $1/2$ which we shall
refer to as a ${\Lambda}_{ij}$ baryon. If the $(Q_iQ_j)$ forms a spin-triplet
(thus flavor-triplet) state the ground-state hadron can be
either a spin-$1/2$ baryon (denoted by ${\Sigma}_{ij}$) or a spin-$3/2$
baryon (denoted by ${\Sigma}_{ij}^{\ast}$ \vspace{0.05in} ).\par
In order to apply the heavy quark spin symmetry to these doubly heavy
baryons, we must realize two different scenarios: $i)$ the
${\lq}{\lq}$outer part" formed by the heavy diquark as a whole and the light
component surrounding it, and $ii)$ the heavy diquark subsystem itself.
Ideally, unbroken heavy quark symmetries imply the mass degeneracy among all
these hadronic states \cite{sav} \cite{whit}. Actually, the chromomagnetic
hyperfine interaction should resolve the degeneracy leading to the mass
level:
\begin{equation}
m({\Sigma}_{ij})\ <\ m({\Lambda}_{ij})\ <\ m({\Sigma}_{ij}^{\ast})
\end{equation}
according to \cite{ito}. Let us remark that these authors apply the hyperfine
splitting to doubly heavy baryons mimicking h-l mesons, that is
they only consider the spin interaction of the heavy diquark with the
light component previously mentioned in $i)$, obtaining
\[ m({\Lambda}_{ij})\ =\ m({\Sigma}_{ij})+\frac{3}{4}{\Delta}m{\mid}_{hyp}\ =\
m({\Sigma}_{ij}^{\ast})-\frac{1}{4}{\Delta}m{\mid}_{hyp} \]
Note that any discussion about the hyperfine splitting {\em inside} the
heavy diquark has apparently been overlooked in \cite{ito} and \cite{sav}. In
fact, the asymptotic behavior of each chromomagnetic interaction is
completely different as stressed in the first part of this work: the former
$i)$ should fall off as any heavy quark mass increases whereas the
latter $ii)$ would finally violate the spin \vspace{0.05in} symmetry.\par
Nevertheless, the inequality (24) should probably hold for $c$, $b$
constituent quarks since the effect of the ${\lq}{\lq}$inner" hyperfine
splitting is suppressed by roughly a factor $(1/2)^4$ with respect
to the ${\lq}{\lq}$outer" hyperfine splitting (a $(1/2)^3$ comes from the
the square of the (Coulombic) wavefunction at the origin and
an additional $1/2$ factor due to $\kappa=c_F{\alpha}_s$ in equation
(9)). Accordingly, observe that the lowest lying state seems likely to be the
${\Sigma}_{bc}$, {\em i.e.} the spin $1/2$ baryon with a diquark which might be
considered as a $J^P=1^+$ \vspace{0.05in} quasi-particle.\par
Finally, let us stress that among all the possible doubly heavy baryons
to be produced at the new generation of hadron colliders, only those with one
$b$ and one $c$ have a real chance to be observed. (This experimental issue
will be discussed in more detail elsewhere \cite{note}). Hereafter, we shall
concentrate on the weakly decaying \vspace{0.1in} spin-1/2
${\Sigma}_{bc}$ baryon.\newline
{\bf ${\Sigma}_{bc}\ \rightarrow\ N_{cc}$ \vspace{0.1in} transitions}\newline
According to our notation, if the heavy diquark
$(bc)$ in the baryon undergoes a semileptonic weak decay yielding another
heavy diquark $(cc)$ (figure 1), the final groundstate baryon can be either a
${\Sigma}_{cc}$ or a ${\Sigma}_{cc}^{\ast}$. Sometimes we shall design
generically both of them as \vspace{0.05in} $N_{cc}$.\par
The quark-level current $J^{\mu}=\overline{c}{\gamma}^{\mu}
(1-{\gamma}_5)b$ responsible for the hadronic transition
${\Sigma}_{bc}{\rightarrow}N_{cc}$  can be written as \cite{whit},
\begin{equation}
J^{\mu}=\hat{\eta}_{12}(v_1{\cdot}v_2)\ \biggl[\ -\frac{1}{2}\
A_{cc}^{\dag\beta}A_{bc\beta}(v_1+v_2)^{\mu}+\frac{1}{2}\
i{\varepsilon}^{\mu\nu\alpha\beta}
A_{cc\nu}^{\dag}A_{bc\alpha}(v_1+v_2)_{\beta}\biggr]
\end{equation}
where $A_{ij}(A_{ij}^{\dag})$ annihilates (creates) an axial-vector
diquark $(ij)$; $\hat{\eta}_{12}(v_1{\cdot}v_2)$ stands for the wavefunction
overlap between the parent and daughter heavy \vspace{0.05in} diquarks.\par
Both initial ${\Sigma}_{bc}$ and final $N_{cc}$ particles belong to
$meson$ supermultiplets of the $SU(6)$ ${\otimes}$ $U(1)$ superflavor
symmetry, as explained in Section 2. Thereby, their well-defined
transformation properties under this symmetry group permits the use of the
trace formalism ensuring a single form factor concerning the overlap of the
light degrees of freedom at leading order approximation. We thus shall write
the overall form factor ${\eta}_{12}$ governing the
hadronic transition factorizing into two pieces \cite{whit}
 \cite{masl}, namely
\begin{equation}
\eta_{12}\ =\ \hat{\eta}_{12}(v_1{\cdot}v_2)\ \xi(v_1{\cdot}v_2)
\end{equation}
At low recoil, the light degrees of freedom only play a marginal role
in the decay, {\em i.e.} $\xi(w\ {\simeq}\ 1)\ {\simeq}\ 1$ as in the analogous
transitions between h-l mesons. Indeed, as shown in ref. \cite{masl},
the light component does not change the overall decay rate of the diquark
though channels the decay into different spin-states according to
some selection rules, as a direct consequence of
QCD in the limit of infinite quark \vspace{0.05in} masses.\par
Since no flavor symmetry
can be applied to h-h systems, neither $\hat{\eta}_{12}$ nor
consequently ${\eta}_{12}$ are normalized to unity at zero recoil but can
be calculated in a model-dependent way as in  \cite{whit} using
Coulomb wavefunctions,
\[ \hat{\eta}_{12}(w=1)\ =\ 8{\chi}^{3/2}\ \ \ ,\ \ \ \biggl(\
{\chi}=\frac{ \hat{m}_{bc}{\alpha}_s(\hat{m}_{bc})\
\hat{m}_{cc}{\alpha}_s(\hat{m}_{cc}) }
{ [\hat{m}_{bc}{\alpha}_s(\hat{m}_{bc})+
\hat{m}_{cc}{\alpha}_s(\hat{m}_{cc})]^2 }\ \biggr) \]
In the numerical evaluation we adopted $m_c=1.5$ GeV, $m_b=4.8$ GeV,
${\alpha}_s(\hat{m}_{bc})\ {\approx}\ 0.3$ and
${\alpha}_s(\hat{m}_{cc})\ {\approx}\ 0.4$. A more exact computation of
$\hat{\eta}_{12}$ using wavefunctions from the Coulomb-plus-linear potential
of reference \cite{quig} provides a very similar numerical value
\footnote{I am indebted to P. Gonz\'alez for technical advice in this
calculation.}, close to \vspace{0.05in} unity.\par
Next, we apply the non-recoil
approximation to the evaluation of rates for the semileptonic decays
as explained in the appendix. One arrives \vspace{0.05in} at
\begin{equation}
\Gamma[{\Sigma}_{bc}{\rightarrow}{\Sigma}_{cc}\ or\ {\Sigma}_{cc}^{\ast}
+{\ell}\overline{\nu}]\ \simeq\
\frac{G_F^2\ m_1^5}{576\ {\pi}^3}\ [5\varphi(x)+2\phi(x)]\ f_{12}^2\
{\mid}V_{cb}{\mid}^2
\end{equation}
Discriminating ${\Sigma}_{cc}$ from ${\Sigma}_{cc}^{\ast}$ in the
final state, one gets readily (see equations (17) and (18) in
reference \cite{masl}).
\begin{equation}
\Gamma[{\Sigma}_{bc}{\rightarrow}{\Sigma}_{cc}+
{\ell}\overline{\nu}]\ \simeq\
\frac{G_F^2\ m_1^5}{576\ {\pi}^3}\
\biggl[\frac{13}{3}\varphi(x)+\frac{4}{3}\phi(x)\biggr]\ f_{12}^2\
{\mid}V_{cb}{\mid}^2
\end{equation}
\begin{equation}
\Gamma[{\Sigma}_{bc}{\rightarrow}{\Sigma}_{cc}^{\ast}+
{\ell}\overline{\nu}]\ \simeq\
\frac{G_F^2\ m_1^5}{576\ {\pi}^3}\
\biggl[\frac{2}{3}\varphi(x)+\frac{2}{3}\phi(x)\biggr]\ f_{12}^2\
{\mid}V_{cb}{\mid}^2
\end{equation}
In Table III we show the numerical results using the last expressions. It is
worth to mention that the relative branching fraction of the semileptonic
decays
${\Sigma}_{bc}{\rightarrow}{\Sigma}_{cc}({\Sigma}_{cc}^{\ast})+
{\ell}\overline{\nu}$ is completely inverted with regard to the decay
$B_c{\rightarrow}{\eta}_c(J/\psi)+\overline{\ell}\nu$. (The underlying
diquark transition is different in each case, $1\ {\rightarrow}\ 1$ in the
former and  $0\ {\rightarrow}\ 0(1)$ in the latter. The groundstate
${\Lambda}_{cc}$ is forbidden by the Pauli exclusion \vspace{0.1in} principle).
\newline
{\bf ${\Sigma}_{bc}\ \rightarrow\ N_b$ \vspace{0.1in} transitions}\newline
Let us assume now that the decay of the ${\Sigma}_{bc}$ proceeds
through the quark decay $c{\rightarrow}s,d$ yielding a final $b$-flavored
baryon. It might be conceivable that the diquark keeps up and the final baryon
could be viewed as made up of a h-l diquark and a light component
\footnote{This may make some sense in particular for an emerging $s$-quark
which could be considered as semi-heavy to some extent; in
fact, such a picture has been recently proposed as an alternative
interpretation for {\em any} baryon containing a single heavy quark
\cite{viet}.}. Another interpretation assumes that
the initial diquark actually breaks up: the $b$ quark becomes a nearly static
source of color whereas one light diquark is formed. We adopt this second
(more standard) possibility henceforth supposing in addition that the $N_b$
lies at the groundstate. Still, the light dressing can be either a spinless
or a spin-1 object. In the following we examine the former case, {\em i.e.}
when
the final baryon is of the \vspace{0.05in} ${\Lambda}_b$-type.
\par
We are addressing here a transition connecting doubly heavy to singly heavy
baryons. In order to take benefit from the compact trace formalism
developed in \cite{geor} for the evaluation of hadronic matrix elements,
we need to recall the fact that both initial and final hadrons belong
to (different) superflavor multiplets under the $SU(6){\otimes}\ U(1)$
spin-flavor symmetry group. Therefore, we shall use the following
$9{\times}4$ matrix representations for the ${\Sigma}_{bc}$ and
${\Lambda}_b$ (anti)baryons,
\begin{center}
\[ {\tilde{\Psi}}_{{\Sigma}_{bc}}(v)\ =\ \frac{1}{\sqrt{6M_{\Sigma}}}\ \left(
\begin{array}{c}  0 \\ 0 \\ u_{\Sigma}^T(v)C{\sigma}^{\mu\alpha}v_{\alpha}
{\gamma}_5
\end{array} \right)\ \ \ ,\ \ \
{\tilde{\Psi}}_{{\Lambda}_b}(v)\ =\ \frac{1}{\sqrt{2M_{\Lambda}}}\
\left( \begin{array}{c} u_{\Lambda}^T(v)C \\ 0 \\ 0 \end{array} \right) \]
\end{center}
where $C$ is the conjugation matrix and the spinors $u_i$ are normalized to
 $\overline{u}_i(v)u_i(v)=2M_i$ satisfying $v{\slash}u_i(v)=u_i(v)$.
The corresponding bra matrices are defined as
\begin{equation}
 \overline{\tilde{\Psi}}(v)\ =\ {\gamma}^0{\tilde{\Psi}}^{\dag}(v)\ \left(
\begin{array}{rrr}
{\gamma}^0\ 0\ 0 \\ 0\ 1\ 0 \\ 0\ 0\ g
\end{array} \right)
\end{equation}
The trace formalism can be employed to compute the hadronic matrix
elements for the ${\Sigma}_{bc}{\rightarrow}N_b$ transition, as
before. There are, however, some crucial differences with respect to the
${\Sigma}_{bc}{\rightarrow}N_{cc}+{\ell}\overline{\nu}$ decay as we shall
\vspace{0.05in} see.
\par
The initial $(bc)$ diquark (with a very large mass ${\simeq}\ m_b+m_c$)
should stand very close to the geometrical center of an spherically symmetric
baryon. Furthermore, the $b$-quark itself would tend to occupy a nearly
central position inside the heavy diquark because of its relative larger
\vspace{0.05in} mass.\par
The mean distance of the $b$ with respect to the center of mass of the $(bc)$
system (and to the center of the baryon as a whole too) is of order
$(\hat{m}_{bc}/m_b)\tilde{r}_0$ where $\tilde{r}_0=3/2{\alpha}_s\hat{m}_{bc}$
denotes the characteristic size of the $(bc)$ bound state,
about twice the Bohr radius of the $B_c$ meson. Thereby, the typical distance
of the $b$-quark to the baryon center is ${\simeq}\ 3/2{\alpha}_sm_b$. On the
other hand, the final heavy quark ({\em i.e.} the $b$) will stay near the
central position of the final baryon as well, oscillating with typical
$\lq\lq$zitterbewegung" frequency $2m_b$ and thus spreading over the
corresponding length scale. Long-distance processes due to soft gluons give
rise to residual momentum exchanges and intermediate quantum states with
energy fluctuations of order $k$, leading also to a typical distance ${\sim}\
1/k{\times}k/m_b=1/m_b$ about the average position of the $b$ quark. Observe
that both of its initial and final oscillation amplitudes are of the same
order (indeed they are $1/m_b$ effects) vanishing in the limit
\vspace{0.05in} $m_b{\rightarrow}\infty$.
\par
Therefore, the decay of the $c$-quark at low-recoil occurs under the influence
of a color field generated by a center-positioned, almost immovable
$b$-quark. (This contrasts with the decay ${\Sigma}_{bc}{\rightarrow}N_{cc}$
where a reassembly of the final {\em heavy} diquark is necessary). Hence, no
form factor relative to the heavy component will be considered in the hadronic
transition amplitude, at least to leading order, or stated in other
words, the wavefunction overlap corresponding to the $b$-quark is
assumed equal to \vspace{0.05in} unity.\par
{}From the above discussion, one can heuristically deduce that solely those
form factors related to the light degrees of freedom will come into play
in the weak current. Accordingly the latter,
$J^{\mu}=\overline{s}{\gamma}^{\mu}(1-{\gamma}_5)c$, acting
on a ${\Sigma}$-type baryon can be written as
\[ J^{\mu}= F^{\mu\nu}(v_1,v_2)\ h_{b}^{\dag}A_{bc\nu} \]
$F^{\mu\nu}$ is a tensor function depending on $v_1$ and $v_2$,
which can be parametrized in general as
\[ F^{\mu\nu}(v_1,v_2)\ =\  a(w)\ g^{\mu\nu}+b(w)\ v_1^{\mu}v_2^{\nu}+
c(w)\ v_2^{\mu}v_2^{\nu}+
d(w)\ i{\varepsilon}^{\mu\nu\alpha\beta}v_{1\alpha}v_{2\beta} \]
where $a$, $b$, $c$, and $d$ are universal functions concerning the light
degrees of freedom. In particular at zero recoil (recall from the appendix
the small range $v_1{\cdot}v_2\ {\leq}\ 1.02$ allowed in this process !) only
the $a$-term gives contribution, all others vanishing since
$v_1^{\nu}A_{bc\nu}(v_1)=0$. The $a(w{\sim}1)$
function absorbs the complexity of the non-perturbative decay dynamics of the
light degrees of freedom near this kinematic point. Let us remark that
it does not coincide with the Isgur-Wise function for transitions among
members of $meson$-type supermultiplets. Nevertheless, $a(w)$ is a true
universal function no matters the flavor of the constituent heavy
\vspace{0.05in} quarks.\par
The hadronic matrix elements near zero recoil can be written as
\begin{equation}
<{\Lambda}_{b}(v){\mid}J^{\mu}{\mid}{\Sigma}_{bc}(v)>\ {\simeq}\
\frac{i\ a(1)}{2\sqrt{3\ m_1m_2}}\ \overline{u}_{\Lambda}(v)\ {\gamma}^{\mu}
{\gamma}_5\ u_{\Sigma}(v)
\end{equation}
Finally, we get in the non-recoil approximation $v_1=v_2$,
\begin{equation}
\Gamma[{\Sigma}_{bc}{\rightarrow}{\Lambda}_{b}+\overline{\ell}\nu]\ {\simeq}\
\frac{G_F^2\ m_1^5}{576\ {\pi}^3}\ [\varphi(x)+\phi(x)]\
f_{12}^2\ {\mid}V_{cs}{\mid}^2
\end{equation}
where \vspace{0.05in} $f_{12}=\sqrt{m_2/m_1}\ a(1)$.\par
In order to have an estimate
of $a(1)$ one may conjecture that its numerical value should be
intermediate between unity ({\em i.e.} the Isgur-Wise function at zero
recoil) and the overlap of the light degrees of freedom corresponding to the
$D{\rightarrow}\ K$ transition (see figure 2). We tentatively adopt the last
wavefunction overlap factor obtained from the ISGW model with the parameters of
Table A.I as an acceptable estimate for $a(1)$. Indeed, note that the presence
of the almost immovable $b$-quark acting as an attractive center of color force
should somehow shrink the size of the light degrees of freedom (especially
compared to the Kaon system where such an static force center does not exist at
all) thereby increasing their wavefunction overlap. However, this effect
should be somewhat cancelled by the fact that the quark-quark interaction is
about half as attractive as in the quark-antiquark case. Our predictions (with
a large uncertainty) are shown in \vspace{0.05in} Table III.
\par
\subsection{Absolute branching fractions}
In order to present our results in a more suitable way from the experimental
viewpoint, we tentatively assume that the full width of the ${\Sigma}_{bc}$
baryon is similar to the full width of the $B_c$ baryon. This assumption can be
somehow justified by the fact that the ${\Lambda}_b$ baryon has similar
lifetime as the $B$ meson \vspace{0.05in} \cite{pdg}.\par
The main objection to this motivated analogy may arise from the fact the
annihilation channel of the meson $B_c{\rightarrow}\ \overline{\ell}\nu$ has no
analogous counterpart in the ${\Lambda}_{bc}$ baryon decay. However, noting
that the expected branching ratio of the $B_c$ into
all leptonic channels is ${\approx}\ 18\%$ \cite{lusi} while
the main part of the decay ($82\%$) corresponds to $\lq\lq$spectator"
diagrams, it remains conceivable that the ${\Sigma}_{bc}$'s lifetime
is not very different from the $B_c$'s \cite{lusi}. Therefore, adopting
$\tau({\Sigma}_{bc})\ {\simeq}\ 0.5$ ps we get the set of branching
fractions shown in Table III (since one would expect
$\tau(B_c)<\tau({\Sigma}_{bc})$ we are underestimating these branching
fractions in some \vspace{0.05in} sort).\par
\subsection{Two-body nonleptonic decay of the ${\Sigma}_{bc}$}
In this Section we shall examine the two-body weak decay
generated by the quark-level process $b{\rightarrow}c\overline{c}s$
\[ {\Sigma}_{bc}\ {\rightarrow}\ {\Xi}_c\ +\ J/\psi \]
where ${\Xi}_c$ denotes a ground-state ${\Lambda}_Q$-type baryon. We have
focused on this exclusive channel yielding a $J/\psi$ since its subsequent
decay into a lepton pair could be useful to provide a tagging signal among
the huge hadronic background at LHC \cite{note}. Proceeding throughout in a
very naive way, the hadronic matrix element
\begin{equation}
<{\Xi}_c\ {J/\psi}\ {\mid}\ J^{\mu}j_{\mu}\ {\mid}\ {\Sigma}_{bc}>
\end{equation}
where $J^{\mu}=\overline{c}{\gamma}^{\mu}(1-{\gamma}_5)b$,
$j^{\mu}=\overline{s}{\gamma}^{\mu}(1-{\gamma}_5)c$ can be Fierz-reordered.
Thus assuming in addition the factorization hypothesis, it can be written as
\begin{equation}
<{\Xi}_c\ {\mid}\ \overline{s}{\gamma}^{\mu}(1-{\gamma}_5)b\ {\mid}\
{\Sigma}_{bc}>\ <J/\psi\ {\mid}\ \overline{c}{\gamma}^{\mu}(1-{\gamma}_5)c\
{\mid}\ 0>
\end{equation}
We next relate this amplitude to the semileptonic
decay at $q^2=m_{J/\psi}^2$ according to \cite{bjor} as we did for the $B_c$,
\[ \frac{\Gamma({\Sigma}_{bc}{\rightarrow}\ {\Xi}_c+J/{\psi})}
{d{\Gamma}/dq^2{\mid}_{q^2=m_{\psi}^2}({\Sigma}_{bc}{\rightarrow}\
N_c+{\ell}\overline{\nu})}\ {\simeq}\ 6{\pi}^2f_{J/\psi}^2s_F\
{\mid}V_{cs}{\mid}^2 \]
where $s_F$ stands for a {\em suppression} factor that we assume to be $4/9$
(the factor 9 comes from color mismatch whereas the enhancing factor $4$
is due to the existence of two final $c$ quarks to combine with the emerging
$\overline{c}$ to yield the $J/\psi$). Note that the denominator represents an
$\lq\lq$unreal" semileptonic process as an artefact with similar initial and
final baryon states as the real physical process in the numerator. Then
we use the following expression (see eq. (A.7)) in the non-recoil approximation
\[ \frac{d{\Gamma}(1{\rightarrow}0)}{dq^2}=
\frac{G_F^2[{\lambda}^{3/2}+12q^2m_1^2{\lambda}^{1/2}]}
{576\ {\pi}^3\ m_1^3}\ f_{12}^2\ {\mid}V_{cb}{\mid}^2 \]
With regard to the overlap factors contained in $f_{12}$ we shall stretch the
arguments used for the semileptonic decay
${\Sigma}_{bc}{\rightarrow}N_b+{\ell}\nu$, whereas this time the
$c$-quark (always remaining within a region of dimension ${\simeq}\ 1/m_c$)
playing the spectator role instead of the $b$. Therefore, we
(tentatively) disregard any form factor due to the rearrangement of the heavy
component, whereas we evaluate the light component overlap
according to the ISGW model like a $\lq\lq$$B{\rightarrow}K$" \vspace{0.05in}
transition.
\par
In table IV we present our predictions for the partial width and branching
fraction of the process. It is interesting to note that the BF
turns out to be of order $1\%$. Of course, such estimate
suffers from many theoretical uncertainties but, however, is in remarkable
accordance with an equivalent calculation by Mannel and Roberts \cite{man}
for the analogous ${\Lambda}_b{\rightarrow}{\Lambda}\ J/\psi$ decay
once taken into account the enhancing factor $4$.
\section{Summary}
\par
We have analyzed the dynamics of the weak decay of hadrons containing
two heavy quarks employing the formalism of the Heavy Quark Effective
Theory to some extent. Firstly we focused on the internal dynamics
of heavy-heavy bound systems. Notice that contrary to heavy-light
systems, there is no flavor symmetry arising from the leading-order
Schr\"{o}dinger-type \vspace{0.05in} Lagrangian.
\par
We showed on the basis of estimates from potential
models that in the range of masses limited to the preasymptotic domain
including $J/\psi$, $B_c$ and $\Upsilon$ states, the Pauli chromomagnetic term
in the effective Lagrangian behaves as a genuine subleading term in a $1/m_Q$
expansion. For heavy quark masses larger than ${\simeq}\ 10$ GeV, the Coulomb
regime will prevail and the spin symmetry will be definitely \vspace{0.05in}
lost.\par
On the other hand, the interaction of a heavy diquark with a light component
in a hadron exhibits a $\lq\lq$superflavor" symmetry arising in the low-energy
limit of the strong interactions, meaning that in the infinite mass limit
both spin and flavor of the color source become irrelevant. Thereby, the
massive quark and diquark fields can be joined into a common $\lq\lq$quark"
supermultiplet as far as their interactions with light degrees of freedom are
concerned. Furthermore, such $\lq\lq$quark" supermultiplet can be combined
with either a light quark or a light diquark field to yield a classification
scheme of hadrons according to the behavior of
the light component. In particular, one can form a $meson$-type multiplet
including singly heavy mesons and doubly heavy (anti)baryons, all of them with
well-defined transformations properties under the superflavor group. All
members
of a hadron supermultiplet, though with different quantum numbers, would show
similar spectroscopy concerning the excitations of the light component. In a
parallel way, one can form two kinds of $baryon$-type multiplets by combination
of the $\lq\lq$quark" supermultiplet with either a spinless or a spin-1
light component. Thus, singly heavy baryons and exotic multiquark states
(doubly heavy diquonia $Q_iQ_j\overline{q}\ \overline{q}$)
would belong to the same \vspace{0.05in} supermultiplet.
\par
In this work we made a phenomenological application of all the above
ideas to the decays of hadrons containing $b$ and $c$ quarks in particular.
Besides, we focused on those exclusive channels leading (at
least) to a final charged lepton. In fact, to be realistic, the presence of
a $J/\psi$ resonance decaying into a charged lepton pair will likely be an
almost necessary condition to disentangle the decay from the huge hadronic
background in hadron colliders (for more details \vspace{0.05in} see
\cite{note}).\par
Firstly, use was made of spin symmetry in the analysis
of some weak decays of the $B_c$ meson. Predictions on partial widths and
branching fractions by means of the non-recoil approximation are shown in
Tables I, II. In general our evaluations are in agreement with other quark
model calculations clearly favoring the final spin-triplet state versus the
spin-singlet state. (However, a significant discrepancy to this general rule
appears in the work of Bagan \vspace{0.05in} {\em et al} using QCD sum rules).
\par
Motivated by this general agreement, we have extended our formalism to the
analysis of some weak decays of the ${\Sigma}_{bc}$ baryon. One ought to
sort out two possible scenarios depending on whether the
decaying quark is either the bottom or the charm \vspace{0.05in} quark.\par
Let us suppose that the $b$ quark undergoes the weak decay yielding a doubly
heavy final baryon ($N_{cc}$) and we look upon it as made of one
heavy diquark along with light degrees of freedom. Then, both parent and
daughter particles belong to $meson$-type multiplets and the trace formalism
based on the superflavor symmetry developed by Georgi {\em et al} \cite{geor}
can be applied to \vspace{0.05in} straightforwardly.\par
In this case, two form factors are required to describe the hadronic
transition at leading order in the effective theory. One of them measures the
overlap of the wavefunctions of the initial and final heavy diquarks, while the
other is simply the Isgur-Wise function, common to transitions among members
of this kind of supermultiplet. Moreover, near zero recoil the light component
should not change the overall decay of the heavy \vspace{0.05in} diquark.\par
Predictions on decay widths and branching
fractions for semileptonic decays adopting the non-recoil approximation
and assuming similar lifetimes for the ${\Sigma}_{bc}$
baryon and the $B_c$ meson can be found in \vspace{0.05in} Table III.\par
Instead, if the transition proceeds through the $c$-quark decay and the final
baryon consists of the surviving $b$-quark and one light diquark, we are
actually confronted to a distinct physical situation. Indeed, the final
hadron belongs to a $baryon$-type supermultiplet whereas the initial
one is of the $meson$-type. We are thus dealing with a transition
between members of {\em different} \vspace{0.05in} supermultiplets.\par
Notice that although the heavy diquark would break up, the $b$-quark would
remain within a small central region of the baryon before and after the
disintegration, in this particular channel. Therefore no form factor measuring
the rearrangement of the heavy component in the baryon comes into play now at
leading order. Instead, we introduced a form factor related to
the light component in analogy to $D{\rightarrow}K$ transitions. Table III
shows our predictions for the partial width and
branching \vspace{0.05in} fraction.\par
Lastly, we have evaluated the expected branching fraction for the
two-body nonleptonic decay of the ${\Sigma}_{bc}$ baryon into
${\Xi}_c$ and $J/\psi$ relying on the factorization hypothesis. Our
results are presented in Table IV.\par
\section*{Acknowledgments}
I thank M. Neubert and S. Narison for useful discussions and  critical
reading. I also thank P. Gonz\'alez for his advice on potential
models. I thank A. Pineda and J. Soto for sending their preprint prior
to publication. Finally, I want to mention P. Eerola and N. Ellis in
particular and the B-physics group of the ATLAS collaboration at the LHC
for their interest and comments on experimental aspects and consequences
of this \vspace{0.05in} work.\par
\newpage
\appendix
\renewcommand{\theequation}{\thesection.\arabic{equation}}
\section*{Appendix}
\section{}
\setcounter{equation}{0}
\vspace{0.05in}
In this paper, the hadronic matrix elements for weak transitions of doubly
heavy hadrons are evaluated employing the non-recoil approximation (the
parent and daughter hadrons are at rest in their common rest frame).
We will retain, nevertheless, relativistic phase space factors in the
calculation of decay widths but leptonic masses will be \vspace{0.05in}
neglected.\par
Throughout, subscripts $1$ and $2$ will
design initial and final quantities. In particular $m_1$, $v_1$, $p_1$ and
$m_2$, $v_2$, $p_2$ stand for the mass, four-velocity and four-momentum of
the initial and final hadron \vspace{0.05in} respectively.\par
The validity of the non-recoil approximation as an estimate is strongly
supported by the fact that the kinematic variable $w=v_1{\cdot}v_2$ is
restricted to values close to unity. To appreciate this, let
us consider a hadron containing two heavy quarks denoted by $Q_i$ and $Q_j$.
Let quark $Q_j$, for example, undergo a weak decay into $Q_k$. At
$q^2=(p_1-p_2)^2=0$,
\begin{equation}
(v_1{\cdot}v_2)_{max}=1\ +\ \frac{(m_1-m_2)^2}{2m_1m_2}\ {\simeq}\
1\ +\ \frac{(m_{Q_j}-m_{Q_k})^2}{2(m_{Q_i}+m_{Q_j})(m_{Q_i}+m_{Q_k})}
\end{equation}
Considering for definiteness the decay $c{\rightarrow}s$ in an initial
hadron containing one bottom quark/antiquark and one charm quark/antiquark,
\[ (bc)_{v_1}\ {\rightarrow}\ (bs)_{v_2}\ :\ \ \ \ \ \ \ \
(v_1{\cdot}v_2)_{max}\ {\simeq}\ 1.02 \]
for the mass values $m_b=4.8$ GeV, $m_c=1.5$ GeV and $m_s=0.3$
\vspace{0.05in} GeV.\par
If the $b$ undergoes the weak decay,
\[ (bc)_{v_1}\ {\rightarrow}\ (cc)_{v_2}\ :\ \ \ \ \ \ \ \
(v_1{\cdot}v_2)_{max}\ {\simeq}\ 1.29 \]
Analogous values should also hold for mesons and baryons with charm and
bottom. Such small ranges of the velocities product should render
the estimates of widths from hadronic amplitudes evaluated at $w=1$ as an
acceptable \vspace{0.05in} first approximation.\par
On the other hand, the spin symmetry exhibited by eq. (7) permits to represent
the covariant spin wavefunctions of doubly heavy bound states in a
similar way as mesons with a single heavy quark, that is \cite{neub} \cite{fal}
\begin{equation}
J^P=0^-:\ \cal{H}(v)=-\sqrt{M}\ P_+{\gamma}_5\ \ \ ;\ \ \
J^P=1^-:\ \ \ \cal{H}(v)=\sqrt{M}\ P_+{\epsilon}\slash
\end{equation}
for a $s$-wave pseudoscalar meson. In the case of a $s$-wave bound state of
two quarks,
\begin{equation}
J^P=0^+:\  \cal{H}(v)=\sqrt{M}\ P_+\ \ \ \ \ \ ;\ \ \
J^P=1^+:\ \ \ \cal{H}(v)=\sqrt{M}\ P_+{\gamma}_5{\epsilon}\slash
\end{equation}
where $P_{\pm}=(1{\pm}v\slash)/2$ are energy \vspace{0.05in} projectors.\par
In this way, the hadronic matrix elements corresponding to
transitions among bound states containing two heavy quarks can be
evaluated according to the well-known trace formalism: \cite{nin}
\begin{equation}
<H_2(v_2){\mid}\overline{Q_k}{\Gamma}Q_j{\mid}H_1(v_1)>\  =\
-\hat{\eta}_{12}(v_1{\cdot}v_2)\ Tr[\overline{\cal{H}}_2(v_2)
{\Gamma}{\cal{H}}_1(v_1)]
\end{equation}
where the initial and final hadronic systems can be scalar, pseudoscalar,
vector or pseudovector states; $\Gamma$ is a combination of
$\gamma$ matrices and $\hat{\eta}_{12}$ plays the role of the dimensionless
Isgur-Wise function for transitions between h-l mesons \cite{neub},
measuring as well the overlap of the initial and final hadronic wavefunctions.
Yet, $\hat{\eta}_{12}(w)$ depends on the type of hadrons
involved in the process and it is not a universal function of the velocity
difference only. On the other hand, unbroken spin symmetry ensures a single
$\hat{\eta}_{12}$ (reduced) form factor for hadronic transitions, irrespective
of $\Gamma$, between different spin-states. Finally, let us stress that
$\hat{\eta}_{12}$ is not absolutely normalized to unity at zero recoil via the
vector current since h-h systems do not belong to a multiplet
of a heavy quark flavour symmetry group (there should be a mismatch between
the parent and daughter hadron wavefunctions even at zero \vspace{0.05in}
recoil).\par
A summary of the relevant hadronic matrix elements in the non-recoil
approximation ($v_1=v_2=v$) for the weak process
$Q_j{\rightarrow}Q_k\ W^{\ast}$ is given by:
\begin{eqnarray}
<{0^P},{\epsilon}_2\ {\mid}\ V^{\mu}\ {\mid}\ {0^P}>\ \ \
& {\simeq} & \ {\pm}\ 2\ \hat{\eta}_{12}\
\sqrt{m_1m_2}\ v^{\mu} \nonumber \\
<{1^P}, {\epsilon}_2\ {\mid}\ A^{\mu}\ {\mid}\ {0^P}>\ \ \
& {\simeq} & \ {\pm}\ 2\ \hat{\eta}_{12}\
\sqrt{m_1m_2}\ {\epsilon}_2^{\ast\mu} \nonumber \\
<{0^P}\ {\mid}\ A^{\mu}\ {\mid}\ {1^P},
{\epsilon}_1>  & {\simeq} & \ {\pm}\ 2\ \hat{\eta}_{12}\
\sqrt{m_1m_2}\ {\epsilon}_1^{\mu} \\
<{1^P},{\epsilon}_2\ {\mid}\ V^{\mu}\ {\mid}\ {1^P},
{\epsilon}_1>  & {\simeq} & \ {\mp}\ 2\ \hat{\eta}_{12}\
\sqrt{m_1m_2}\ ({\epsilon}_1{\cdot}\ {\epsilon}_2^{\ast})\ v^{\mu}
\nonumber \\
<{1^P},{\epsilon}_2\ {\mid}\ A^{\mu}\ {\mid}\ {1^P},
{\epsilon}_1>  & {\simeq} & \ {\mp}\ 2\ \hat{\eta}_{12}\
\sqrt{m_1m_2}\ i\ {\varepsilon}^{\mu\nu\alpha\beta}\ v_{1\nu}\
{\epsilon}_{1\alpha}\ {\epsilon}_{2\beta}^{\ast} \nonumber
\end{eqnarray}
where the upper(lower) sign applies when $P={\mp}$ respectively;
the vector and axial-vector currents are
$V^{\mu}=\overline{Q}_k{\gamma}^{\mu}Q_j$ and
$A^{\mu}=\overline{Q}_k{\gamma}^{\mu}{\gamma}_5Q_j$.
Upon contraction with the leptonic current
$\overline{\ell}{\gamma}_{\mu}(1-{\gamma}_5)\nu$, squaring, summing
and averaging over polarization states whenever needed, the resulting
differential widths for each massless lepton species and charge mode are
\begin{equation}
\frac{d\Gamma}{dq^2}(0^P{\rightarrow}0^P){\mid}_{vector}\ {\simeq}\
\frac{d\Gamma}{dq^2}(1^P{\rightarrow}1^P){\mid}_{vector}\ {\simeq}\
\frac{G_F^2{\lambda}^{3/2}}{192\ {\pi}^3\ m_1^3}
\ f_{12}^2\ {\mid}V_{jk}{\mid}^2
\end{equation}
\begin{equation}
\frac{d\Gamma}{dq^2}(1^P{\rightarrow}0^P){\mid}_{axial}\ {\simeq}\
\frac{1}{3}\frac{d\Gamma}{dq^2}(0^P{\rightarrow}1^P){\mid}_{axial}\ {\simeq}\
\frac{1}{2}\frac{d\Gamma}{dq^2}(1^P{\rightarrow}1^P){\mid}_{axial}\ {\simeq}\
\frac{G_F^2({\lambda}^{3/2}+12q^2m_1^2{\lambda}^{1/2})}{576\ {\pi}^3\ m_1^3}
\ f_{12}^2\ {\mid}V_{jk}{\mid}^2
\end{equation}
where $\lambda{\equiv}\lambda(m_1^2,m_2^2,q^2)$ denotes the K\"{a}llen
function, $V_{jk}$ is the KM mixing matrix element involved in the quark's
decay. The (single) hadronic form factor $f_{12}$ at the non-recoil
approximation is related to $\hat{\eta}_{12}$ by:
\begin{equation}
f_{12}=\sqrt{\frac{m_2}{m_1}}\ \hat{\eta}_{12}(w=1)
\end{equation}
assumed nearly constant over the whole available $q^2$ range. This introduces
an uncertainty lying within the framework of the approximations
made in this work. Using the ISGW model \cite{igsw} at $q_{max}^2$ we
may \vspace{0.05in} write
\begin{equation}
\hat{\eta}_{12}\ =\
\biggl(\ \frac{2{\beta}_1{\beta}_2}{{\beta}_1^2+{\beta}_2^2}\ \biggr)^{3/2}
\end{equation}
Those values of the ${\beta}_i$ parameter of interest in our
work are collected in table \vspace{0.05in} A.I.\par
Integrating over $q^2$ between $0$ and $q_{max}^2=(m_1-m_2)^2$,
\begin{equation}
\Gamma(0^P{\rightarrow}0^P){\mid}_{vector}\ \simeq\
\Gamma(1^P{\rightarrow}1^P){\mid}_{vector}\ \simeq\
\frac{G_F^2\ m_1^5}{192\ {\pi}^3}\ \varphi(x)\ f_{12}^2\ {\mid}V_{jk}{\mid}^2
\end{equation}
\begin{equation}
\frac{1}{3}\ \Gamma(0^P{\rightarrow}1^P){\mid}_{axial}\ \simeq\
\Gamma(1^P{\rightarrow}0^P){\mid}_{axial}\ \simeq\
\frac{1}{2}\ \Gamma(1^P{\rightarrow}1^P){\mid}_{axial}\ \simeq\
\frac{G_F^2\ m_1^5}{576\ {\pi}^3}\ \ [\varphi(x)+\phi(x)]\ f_{12}^2\
{\mid}V_{jk}{\mid}^2\
\end{equation}
where $\varphi$ and $\phi$ are phase space factors depending on the mass
ratio $x=m_2/m_1$ (neglecting lepton masses). Their expressions are: \cite{mas}
\begin{eqnarray}
\varphi(x) & = & \frac{1}{4}\ [(1-x^4)(1-8x^2+x^4)-24\ x^4\ln{x}] \nonumber\\
   \phi(x) & = & 2\ (1-x^2)\ [(1+x^2)^2+8x^2]+24\ (1+x^2)\ x^2\ln{x}  \nonumber
\end{eqnarray}
In the limit $x{\rightarrow}1$, corresponding {\em e.g.} to the
Shifman-Voloshin
condition $(m_1-m_2)^2<<(m_1+m_2)^2$, {\em i.e.} small four-momentum transfer
compared to the hadronic masses, phase space factors behave as:
\begin{equation}
\frac{1}{2}\ \phi(x)\ {\simeq}\ \varphi(x)\ {\rightarrow}\ \frac{16}{5}\
(1-x)^5
\end{equation}
In this extreme non-relativistic limit,
\begin{equation}
\Gamma(0^P{\rightarrow}0^P){\mid}_{vector}\ {\simeq}\
\Gamma(1^P{\rightarrow}1^P){\mid}_{vector}\ {\simeq}\
\frac{1}{3}\ \Gamma(0^P{\rightarrow}1^P){\mid}_{axial}\ {\simeq}\
\Gamma(1^P{\rightarrow}0^P){\mid}_{axial}\ {\simeq}\
\frac{1}{2}\ \Gamma(1^P{\rightarrow}1^P){\mid}_{axial}
\end{equation}
\[
{\simeq}\ \frac{G_F^2\ (m_1-m_2)^5}{60\ {\pi}^3}\ f_{12}^2\ {\mid}V_{jk}
{\mid}^2
\]
Assuming in addition that the spin-singlet and spin-triplet states are
perfectly mass degenerate, the spin-rule yielding a factor $3$ in favor
of final vector states arises from the above expressions at once. In reality,
however, broken spin symmetry may spoil such simple expectations mainly
because of the different available phase space in each case.

\newpage

\vspace{-3cm}

{\bf Figure \vspace{0.2in} Captions:}
\begin{itemize}
\item[{\bf Figure 1)}] Graphs representing the weak decay of doubly heavy
hadrons. For the weak interaction, the graphs may be interpreted as
Feynman diagrams. In the case of the baryon the light quark should act as
spectator at low recoil, whereas the diquark can be considered
as a decaying \vspace{0.3in} quasi-particle.\newline
\item[{\bf Figure 2)}] A visualization of the hadronic transitions:
a) ${\Sigma}_{bc}{\rightarrow}N_{cc}$, b) ${\Sigma}_{bc}{\rightarrow}N_{b}$,
c) $D{\rightarrow}K$. At  zero recoil, the overlap factor of the light degrees
of freedom in the case b) should be somehow intermediate between a) and c).
\end{itemize}
\newpage

\vspace{-3cm}

\begin{table}
{\bf Table I.a).} Partial widths of the semileptonic decays of
$B_c$ ($\overline{b}c$) in units $10^{-15}$ GeV. The relative branching
fraction refers to the final states with a vector meson with respect to those
with a pseudoscalar meson. The same input values of the parameters of the ISGW
model, $V_{cb}=0.046$ etc as in ref. \cite{lusi} have been used to
facilitate the \vspace{0.1in} comparison.
\begin{center}
\begin{tabular}{||l|c|c|c|c|c||}        \hline
Mode  & this work & \cite{lusi} & \cite{kis} & \cite{chan} & \cite{nari}  \\
\hline
$B_c{\rightarrow}\ {\eta}_c+\overline{\ell}{\nu}$ & $10$ & $10.6$ & $15$
& $14.2$ & $9.4$  \\
\hline
$B_c{\rightarrow}\ J/\psi+\overline{\ell}\nu$ & $42$ & $38.5$ & $44$ & $34.4$ &
$12.2$  \\
\hline
relative $BF$ & $4.2$ & $3.6$ & $3$ & $2.4$ & $1.3$  \\
\hline\hline
$B_c{\rightarrow}\ B_s+\overline{\ell}\nu$ & $18$ & $16.4$ & $-$ & $26.6$
& $14.2$  \\
\hline
$B_c{\rightarrow}\ B_s^{\ast}+\overline{\ell}\nu$ & $43$ & $40.9$ & $-$ & $44$
& $14.2$  \\
\hline
relative $BF$ & $2.4$ & $2.5$ & $-$ & $1.7$ & $1$  \\
\hline
\end{tabular}
\end{center}
\vspace{0.1in}
\end{table}

\begin{table}
{\bf Table I.b).} Branching fractions from our results in Table I.a) using
a lifetime of the $B_c$ equal to $0.5$ ps \vspace{0.1in} \cite{lusi}.
\begin{center}
\begin{tabular}{||l|c||}        \hline
Mode  & $BF(\%)$ \\
\hline
$B_c{\rightarrow}\ {\eta}_c+\overline{\ell}\nu$ & $0.75$ \\
\hline
$B_c{\rightarrow}\ J/\psi+\overline{\ell}\nu$ & $3.15$ \\
\hline\hline
$B_c{\rightarrow}\ B_s+\overline{\ell}\nu$ & $1.35$ \\
\hline
$B_c{\rightarrow}\ B_s^{\ast}+\overline{\ell}\nu$ & $3.22$ \\
\hline
\end{tabular}
\end{center}
\vspace{0.1in}
\end{table}

\begin{table}
{\bf Table II.a).} Partial widths of nonleptonic decays of the $B_c$ in
units $10^{-15}$ GeV, where we have used the same input
values as in reference \vspace{0.1in} \cite{lusi}.
\begin{center}
\begin{tabular}{||l|c|c|c||}        \hline
Mode  & this work & \cite{lusi} & \cite{chan} \\
\hline
$B_c{\rightarrow}\ {\eta}_c+\pi$ & $2.4$ & $2.1$ & $3.29$ \\
\hline
$B_c{\rightarrow}\ J/{\psi}+\pi$ & $2.3$ & $2.2$ & $3.14$ \\
\hline
relative $BF$ & ${\sim}1$ & ${\sim}1$ & ${\sim1}$ \\
\hline
\end{tabular}
\end{center}
\vspace{0.1in}
\end{table}

\begin{table}
{\bf Table II.b).} Branching fractions from our results in Table II.a)
using a lifetime of the $B_c$ equal to $0.5$ ps \vspace{0.1in} \cite{lusi}.
\begin{center}
\begin{tabular}{||l|c||}        \hline
Mode  & $BF(\%)$ \\
\hline
$B_c{\rightarrow}\ {\eta}_c+\pi$ & $0.18$ \\
\hline
$B_c{\rightarrow}\ J/{\psi}+\pi$ & $0.17$ \\
\hline
\end{tabular}
\end{center}
\vspace{0.1in}
\end{table}

\begin{table}
{\bf Table III.} Partial widths of semileptonic decays of the
${\Sigma}_{bc}$ baryon ($bcq$) in units $10^{-15}$ GeV. We have used
$M_{{\Sigma}_{bc}}=6.93$ GeV, $M_{{\Sigma}_{cc}}=3.63$ GeV,
$M_{{\Sigma}_{cc}^{\ast}}=3.73$ GeV, $M_{{\Lambda}_b}=6.0$ GeV \cite{nari}.
For the estimates of the absolute BF we set the ${\Sigma}_{bc}$ lifetime
equal to $0.5$ \vspace{0.1in} ps.\newline
\begin{center}
\begin{tabular}{||l|c|c||}        \hline
Mode  & width & BF$(\%)$\\
\hline
${\Sigma}_{bc}{\rightarrow}\ {\Sigma}_{cc}+{\ell}\overline{\nu}$ & $40$
& $3.0$\\
\hline
${\Sigma}_{bc}{\rightarrow}\ {\Sigma}_{cc}^{\ast}+{\ell}\overline{\nu}$ & $12$
& $0.9$ \\
\hline\hline
${\Sigma}_{bc}{\rightarrow}\ {\Lambda}_{b}+\overline{\ell}\nu$ & $34$ & $2.6$
\\
\hline
\end{tabular}
\end{center}
\end{table}
\newpage

\begin{table}
{\bf Table IV).} Width (in units of $10^{-15}$ GeV) and branching fraction
of a two-body nonleptonic decay of the ${\Sigma}_{bc}$ assuming its
lifetime equal to $0.5$ ps, $M_{{\Xi}_c}=2.47$ \cite{pdg} GeV and
$f_{J/\psi}=0.385$ GeV \vspace{0.2in} \cite{lusi}.
\begin{center}
\begin{tabular}{||l|c|c||}        \hline
Mode  & width & $BF(\%)$ \\
\hline
${\Sigma}_{bc}{\rightarrow}\ {\Xi}_{c}+J/\psi$ & $10$ & $0.8$ \\
\hline
\end{tabular}
\end{center}
\vspace{0.1in}
\end{table}

\begin{table}
{\bf Table A.I.} Values in GeV of the ${\beta}_i$ parameter in the ISGW model
\cite{igsw} needed for the evaluation of several wavefunction overlaps at zero
recoil used in this \vspace{0.1in} work.\newline
\begin{center}
\begin{tabular}{||l|c|c|c|c|c|c||}        \hline
$i=$  & $K$ & $D$ & $B$ & $B_s$ & $J/\psi$ & $B_c$ \\
\hline
${\beta}_i$  & $0.31$ & $0.39$ & $0.41$ & $0.51$ & $0.66$ & $0.82$ \\
\hline
\end{tabular}
\end{center}
\end{table}

\end{document}